# Two-dimensional solitons at interfaces between binary superlattices and homogeneous lattices

M. Heinrich,[1] Y. V. Kartashov,[2] L. P. R. Ramirez,[1] A. Szameit,[3] F. Dreisow,[1] R. Keil,[1] S. Nolte,[1] A. Tünnermann,[1] V. A. Vysloukh,[2] and L. Torner[2]

[1]Institute of Applied Physics, Friedrich-Schiller-University Jena, Max-Wien-Platz 1, 07743 Jena, Germany

[2]ICFO-Institut de Ciencies Fotoniques, and Universitat Politecnica de Catalunya, Mediterranean Technology Park, 08860 Castelldefels (Barcelona), Spain

[3]Physics Department and Solid State Institute, Technion, 32000 Haifa, Israel

We report on the experimental observation of two-dimensional surface solitons residing at the interface between a homogeneous square lattice and a superlattice that consists of alternating "deep" and "shallow" waveguides. By exciting single waveguides in the first row of the superlattice, we show that solitons centered on deep sites require much lower powers than their respective counterparts centered on shallow sites. Despite the fact that the average refractive index of the superlattice waveguides is equal to the refractive index of the homogeneous lattice, the interface results in clearly asymmetric output patterns.

## I. Introduction

Inhomogeneous transverse refractive index landscapes provide a variety of possibilities for the control of light propagation. Of particular interest are transversely periodic systems, so-called lattices, where diffraction depends on a variety of parameters such as the depth and frequency of the refractive index modulation, and the propagation angle. In the presence of nonlinearity, such periodic refractive index landscapes support lattice solitons [1,2]. Lattice solitons are a rich phenomenon, comprising amongst others multihump states; solitons carrying vorticity; states with different symmetries along different axes of underlying lattice, and multi-component solitons. The properties and domains of existence of lattice solitons are dictated to a large extent by the internal structure of the lattice. In this respect particularly interesting are lattices with nontrivial shapes, such as superlattices with binary unit cells that provide the possibility to engineer a mini-gap within the first propagation band [3,4]. The properties of spatial gap solitons supported by superlattices were analyzed in [4], gap solitons in Bragg gratings with a harmonic superlattice were addressed in [5], whereas the experimental observation of one-dimensional superlattice gap solitons was conducted in [6]. A number of interesting linear effects, such as Zener tunneling and Bloch-Zener oscillations have also been reported in superlattices [7,8].

On the other hand, truncation of otherwise periodic structures creates an interface between uniform and periodic media. Such interfaces can support lattice surface solitons [9]. Current technologies allow fabrication or optical induction of truncated lattices where surface solitons can be observed at moderate power levels [10]. Surface lattice solitons may exist not only in focusing [9,10], but also in defocusing materials [11], as was demonstrated experimentally [12,13]. One-dimensional surface solitons were analyzed theoretically and observed experimentally in a number of settings and for a variety of truncated lattices, including chirped lattices [14], lattices of Kronig-Penney type [15], and superlattices [16-18]. Also, a variety of linear phenomena was demonstrated in such structures [19-21]. Two-dimensional truncated periodic lattices are also capable of supporting surface solitons. Compared to one-dimensional settings, such interfaces may support much more complicated solitons such as two-dimensional multipoles [22], vortex solitons [23], soliton arrays [24,25], or asymmetric surface states [26,27]. Experimentally, two-dimensional surface solitons at the simplest interfaces between lattice and uniform medium were observed in [28,29]. Surface solitons may reside on defect channels or at two-dimensional interfaces with complex shapes [30,31]. Finally, even fully three-dimensional entities may exist in periodic media such as three-dimensional light bullets may propagating along the interface of periodic materials [32,33] or surface solitons in three dimensions [34].

A particularly interesting class of interfaces occurs between two different periodic structures [35]. In the simplest case such an interface can be formed by two lattices characterized by different refractive index modulation depths or periods. Solitons at such interfaces were analyzed and observed in both one- and two-dimensional settings [36-40]. However, the properties of surface solitons at the interfaces between lattices with different unit cells, e.g. between a homogeneous lattice and superlattice, have not been addressed to this date. In this article, we analyze theoretically and observe experimentally solitons at the interface between homogeneous lattices and superlattices that consist of alternating deep and shallow waveguides fabricated by femtosecond-laser direct inscription [41]. We demonstrate that the thresholds for soliton formation as well as the symmetry of the output patterns strongly depend on whether deep or shallow guides are excited in the first row of superlattice.

## II. The model

We describe the dynamics of beam propagation at the interface between superlattice and homogeneous lattice with the nonlinear Schrödinger equation for the dimensionless field amplitude $q$ assuming CW illumination:

$$i\frac{\partial q}{\partial \xi} = -\frac{1}{2}\left(\frac{\partial^2 q}{\partial \eta^2} + \frac{\partial^2 q}{\partial \zeta^2}\right) - |q|^2 q - R(\eta,\zeta)q. \qquad (1)$$

Here $\eta, \zeta$ and $\xi$ are the transverse and longitudinal coordinates normalized to the characteristic transverse scale and diffraction length, respectively. The refractive index distribution in our structure is described by the function $R(\eta,\zeta)$ that is represented by a sum of Gaussian functions $G(\eta_k,\zeta_k) = p_k \exp[-(\eta - \eta_k)^2/w_\eta^2 - (\zeta - \zeta_k)^2/w_\zeta^2]$ accounting for the elliptical shapes of individual waveguides with widths $(w_\eta, w_\zeta)$ centered around the coordinates $(\eta_k,\zeta_k)$ (see Fig. 1 for a schematic sketch of the lattices). The depth $p_k$ of the refractive index modulation depends on whether the waveguide belongs to the "deep" ($p_k = p_d$) or "shallow" ($p_k = p_s$) sublattice of the superlattice, or to the homogeneous lattice ($p_k = p_h$) that occupies the region below the diagonal of the structure. In the following, we consider the situation of an equal mean refractive index in both homogeneous lattice and superlattice $[(p_d + p_s)/2 = p_h]$. The period of the homogeneous lattice $d$ is equal to the vertical and horizontal spacing of sites in the superlattice, so that the separation between waveguides at both sides of the interface is identical. In accordance with the experimental parameters, our model lattice contains 85 waveguides, the period is set to $d = 6.4$ (corresponding to an actual separation of 64 $\mu$m). We chose modulation depths of $p_s = 2.86$, $p_h = 2.93$, and $p_d = 3.00$. Notice, that a depth of $p_{s,h,d} = 3$ is equivalent to an actual refractive index modulation depth of $\delta n \sim 3.3 \times 10^{-4}$. We excite waveguides located in the first row of the superlattice (see sites marked by a red circumference in Fig. 1). In the following, we refer to the case of an excited deep waveguide as $D$-lattice interface (Fig. 1, left) and an excited shallow waveguide as $S$-lattice interface (Fig. 1, right), respectively. For the characterization of soliton solutions we introduce the total energy flow $U$ which is a conserved quantity of Eq. (1), and the integral width $W$ of a soliton:

$$U = \int\int_{-\infty}^{\infty} |q|^2 \, d\eta d\zeta,$$
$$W = \int\int_{-\infty}^{\infty} (2/U)|q|^2 (\eta^2 + \zeta^2)^{1/2} d\eta d\zeta. \qquad (2)$$

### III. Numerical results

We search for stationary solutions of Eq. (1) centered around a site in the first row of $D$ or $S$ superlattices in the form $q = w(\eta, \zeta) \exp(ib\xi)$, where $b$ is the propagation constant. All solutions of that type exist above a cutoff propagation constant $b_{co}$ and for energy flows exceeding a certain threshold value $U_{th}$. For both types of interfaces ($D$-lattice – homogeneous lattice, or $S$-lattice – homogeneous lattice) the energy flow is a nonmonotonic function of propagation constant: $U$ decreases with decreasing $b$, then reaches the threshold value in a global minimum, and finally grows again as $b \to b_{co}$ [Fig. 2(a)]. The width of the soliton monotonically decreases with an increasing propagation constant $b$, while the dependence $W(U)$ is more complicated [Fig. 2(b)]. Representative profiles of interface solitons are shown in Fig. 3: While for large values of $b$ and for large peak amplitudes the solitons at both types of excitations contract to a singe lattice site [Figs. 3(b) and 3(d)], they dramatically expand across the lattice close to the cutoff [Figs. 3(a) and 3(c)]. Notice however, that this expansion is much stronger in the case of solitons at the $D$-lattice interface [Fig. 3(a)]. Solitons at the $S$-lattice interface tend to stretch only across a few neighboring channels as $b \to b_{co}$ [Fig. 3(b)]. This behavior is also evident from Fig. 2(b), which shows that width of solitons at the $D$-lattice interface may take on much larger values. Despite the fact that the average refractive indices are equal on both sides of the interface, the stationary patterns are clearly asymmetric: Light tends to penetrate deeper into the superlattice region than into the homogeneous lattice. Due to the large but finite number of waveguides in both model and experimental arrays, the cutoff for the existence of solitons at the $D$-lattice interface is slightly lower than in the case of the $S$-lattice. Importantly, the energy flow threshold for existence of surface solitons considerably differs for the two types of excitations. Consequently, it is much easier to excite surface solitons at the $D$-lattice interface where $U_{th} \approx 0.546$ than at the interface of the $S$-lattices where $U_{th} \approx 0.921$. The discrepancy in thresholds becomes even more pronounced if the detuning $p_d - p_s$ is increased while keeping the average refractive index $(p_d + p_s)/2$ fixed. According to both linear stability analysis and the Vakhitov-Kolokolov criterion [42] surface solitons of our system are stable when they belong to the branch of $U(b)$, where $dU/db > 0$, but exponentially unstable otherwise.

### IV. Experimental results

Our experiments were conducted in waveguide arrays resembling the model systems discussed above. The arrays were fabricated in fused silica using the femtosecond-laser direct writing technique [41]. The sample length was 105 mm, specific fabrication parameters are discussed in [30]. A Ti:Sapphire laser system (Spectra Physics Tsunami Spitfire), delivering 200 fs pulses at a wavelength of 800 nm with a repetition rate of 1 kHz was used to excite the lattice sites under investigation with a 2.5x microscope objective (NA=0.075). The intensity distributions at the sample's output facet were imaged onto a CCD camera with a 4x microscope objective (NA=0.10).

Figure 4 shows the simulated (top row) and observed (bottom row) output intensity distributions at specific injection peak powers for excitation of the central waveguide in the $D$-lattice. In column (a), the linear diffraction patterns are presented. Here, light spreads equally well into both the superlattice and the homogeneous region; the asymmetry of the overall pattern with respect to the interface is evident. At an intermediate power level of 1.06 MW the diffractive broadening of the pattern is already slightly decreased, [column (b)]. Notably, the fraction of light propagating in the superlattice domain increases, as the nonlinear contributions detune the effective index of the excited guide further from that of the homogeneous region. The contraction progresses monotonously [see patterns at 1.29 MW in column (c)], until at 2.24 MW light remains trapped in the excited guide. The observed nonlinear patterns feature a moderate background due to conical emissions from the pulse slopes, since pulsed illumination was used in the experiments [43].

In Figure 5 the output intensity distributions in the $S$-lattice are shown, the arrangement of figures corresponds to Fig. 4. Although in the linear case [column (a)], light again spreads into both regions, the asymmetry of the overall pattern is slightly more distinct than in the case of the $D$-lattice interface. At the onset of nonlinear propagation around 1.18 MW, light is notably drawn into the homogeneous region, since the effective refractive index of the excited guide becomes equal to that of the homogeneous domain [column (b)]. Around 1.53 MW, this effect is reversed when the effective index reaches the value of the surrounding deep guides of the superlattice. Furthermore, the overall pattern starts to contract [column (c)]. However, as predicted by theory, substantially higher powers are re-quired to achieve complete localization. Hence, at the highest experimentally accessible excitation peak power of 3.18 MW the corresponding output pattern still comprises also the adjacent lattice sites on both sides of the interface [column (d)]. Note that the background in all nonlinear measurements for the $S$-lattice interface is much more pronounced than at the $D$-lattice interface: when the index detuning between the excited shallow guide and the adjacent guides of the homogeneous region vanishes, coupling reaches a maximum. Thus, even at higher powers a significant fraction of light from the pulse slopes will be radiated into the homogeneous region. Similarly, additional losses occur due to index matching to the surrounding deep guides.

Note that while an exact quantitative comparison between simulations and measurements is difficult due to the background in the experiments, the results are in good qualitative agreement and clearly illustrate the underlying physical characteristics of two-dimensional superlattice interface solitons.

## V. Conclusions

In conclusion, we studied theoretically and observed experimentally the formation of solitons residing at the interface between a two-dimensional binary superlattice and a homogenous lattice. We showed numerically that solitons centered on deep sites feature significantly lower power thresholds than their respective counterparts centered on shallow sites. Despite the equal mean refractive indices of both regions, the intensity distributions are always asymmetric with respect to the interface since light tends to penetrate deeper into the superlattice domain. At intermediate power levels, nonlinear index matching causes solitons on shallow sites to shift their barycenter towards the homogeneous region.

## Acknowledgements

The authors gratefully acknowledge support by the Deutsche Forschungsgemeinschaft (Research Unit 532 and Leibniz program), the German Academy of Science Leopoldina (grant LPDS 2009-13), and the Government of Spain (Ramon-y-Cajal program).

# Figure captions

Figure 1 (color online).    A schematic sketch of the interface between $D$-lattice and homogeneous lattice (left) and interface between $S$-lattice and homogeneous lattice (right). The dark and light gray sites correspond to the deep and shallow waveguides of the superlattice, while the medium gray sites represent the homogeneous lattice. In both lattices the excitation was placed in central waveguides.

Figure 2.    (a) Energy flow versus propagation constant and (b) width versus energy flow for solitons at the $D$-lattice and $S$-lattice interface respectively. The circles correspond to the soliton profiles shown in Fig. 3. All quantities are plotted in normalized units.

Figure 3 (color online).    Field normalized modulus distributions for solitons at the $D$-lattice interface at (a) $b=0.314$ and (b) $b=0.433$, and for solitons at the $S$-lattice interface at (c) $b=0.327$ and (d) $b=0.414$. The white dashed lines indicate the interface. The bars in panel (d) indicate transverse scales that are identical for all panels in this figure.

Figure 4 (color online).    Comparison of the normalized output intensity distributions for an excitation of a surface waveguide at the $D$-lattice interface. Top row - experiment, bottom row - theory. The input peak power is (a) 0.12 MW, (b) 1.06 MW, (c) 1.29 MW, and (d) 2.24 MW. The bars in (d), lower panel, indicate transverse scales that are identical for all panels in this figure.

Figure 5 (color online).    Comparison of the normalized output intensity distributions for an excitation of a surface waveguide at the $S$-lattice interface. Top row - experiment, bottom row - theory. The input peak power is (a) 0.12 MW, (b) 1.18 MW, (c) 1.53 MW, and (d) 3.18 MW. The bars in (d), lower panel, indicate transverse scales that are identical for all panels in this figure.

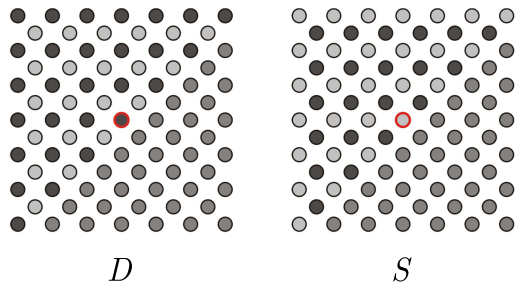

Figure 1 (color online). A schematic sketch of the interface between $D$-lattice and homogeneous lattice (left) and interface between $S$-lattice and homogeneous lattice (right). The dark and light gray sites correspond to the deep and shallow waveguides of the superlattice, while the medium gray sites represent the homogeneous lattice. In both lattices the excitation was placed in central waveguides.

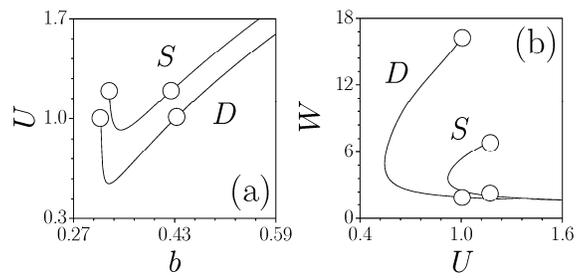

Figure 2.  (a) Energy flow versus propagation constant and (b) width versus energy flow for solitons at the $D$-lattice and $S$-lattice interface respectively. The circles correspond to the soliton profiles shown in Fig. 3. All quantities are plotted in arbitrary dimensionless units.

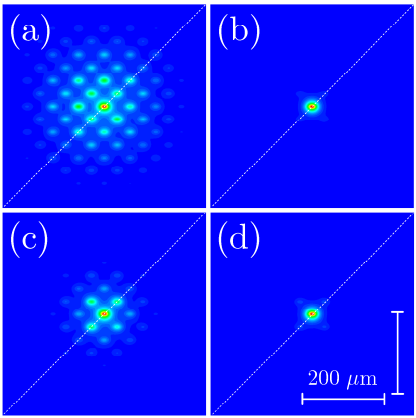

Figure 3 (color online). Field modulus distributions for solitons at the $D$-lattice interface at (a) $b = 0.314$ and (b) $b = 0.433$, and for solitons at the $S$-lattice interface at (c) $b = 0.327$ and (d) $b = 0.414$. The white dashed lines indicate the interface. The bars in panel (d) indicate transverse scales that are identical for all panels in this figure. Field modulus is plotted in dimensionless units.

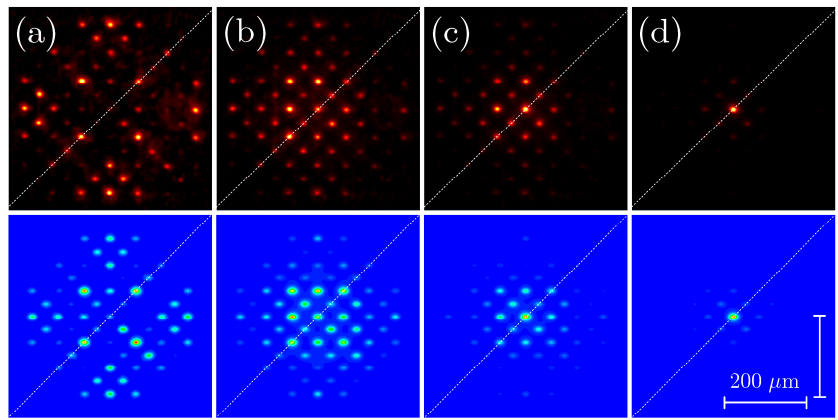

Figure 4 (color online). Comparison of the output intensity distributions for an excitation of a surface waveguide at the $D$-lattice interface. Top row - experiment, bottom row - theory. The input peak power is (a) 0.12 MW, (b) 1.06 MW, (c) 1.29 MW, and (d) 2.24 MW. The bars in (d), lower panel, indicate transverse scales that are identical for all panels in this figure. Intensity is plotted in dimensionless units.

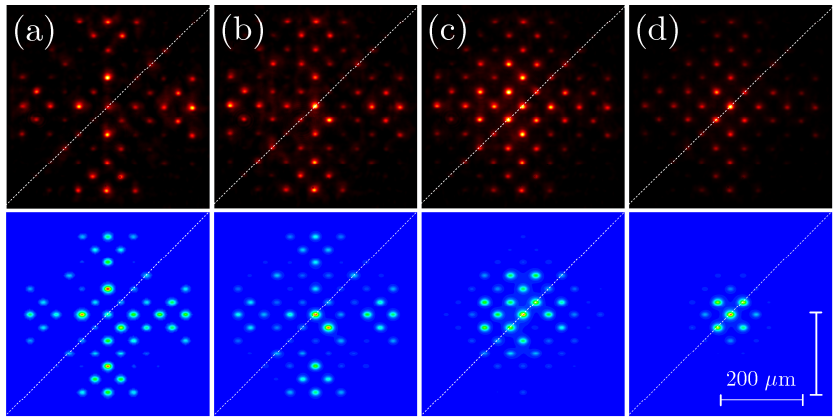

Figure 5 (color online). Comparison of the output intensity distributions for an excitation of a surface waveguide at the $S$-lattice interface. Top row - experiment, bottom row - theory. The input peak power is (a) 0.12 MW, (b) 1.18 MW, (c) 1.53 MW, and (d) 3.18 MW. The bars in (d), lower panel, indicate transverse scales that are identical for all panels in this figure. Intensity is plotted in dimensionless units.

16